\documentclass[nofootinbib,a4paper,aps,pra,showpacs,preprintnumbers,twocolumn]{revtex4-1}
\usepackage{cmap}
\usepackage{graphics,graphicx,epsfig}
\usepackage{epstopdf}
\usepackage[centertags]{amsmath}
\usepackage{amsfonts}
\usepackage{amssymb}
\usepackage[cp1251]{inputenc}
\usepackage[english]{babel}
\inputencoding{cp1251}
\usepackage{graphicx}
\usepackage{amsthm,color}
\usepackage{euscript}
\DeclareGraphicsExtensions{.pdf,.png,.jpg,.eps}
\begin{document}
\def\BY{\begin{eqnarray}}
\def\EY{\end{eqnarray}}
\def\L{\label}
\def\nn{\nonumber}
\def\ds{\displaystyle}
\def\o{\overline}
\def\({\left (}
\def\){\right )}
\def\[{\left [}
\def\]{\right]}
\def\<{\langle}
\def\>{\rangle}
\def\h{\hat}
\def\td{\tilde}
\def\r{\vec{r}}
\def\ro{\vec{\rho}}
\def\h{\hat}
\def\v{\vec}
\title{Conversion and storage of modes with orbital angular momentum in quantum memory scheme}
\author{Vashukevich E.A., Golubeva T.Yu., Golubev Yu.M.}

\address{Saint Petersburg State University, Universitetskaya emb. 7/9, St. Petersburg, 199034 Russia}
\begin{abstract}
The paper studies the Raman quantum memory protocol as applied to quantum light with orbital angular momentum. The memory protocol is implemented on an ensemble of three-level cold atoms with the $\Lambda$- configuration of energy levels. The possibility of storing quantum statistics of light with an orbital momentum is analysed in the case when the driving field could be treated as a plane wave. The efficiency analysis shows that examined storage/retrieval processes do not cause the efficiency decreasing compared with the spatial multimode memory protocol considered in \cite{Golubeva2012}. We also present an effective transformation of the orbital angular momentum of a quantum field on a memory cell using the driving field with orbital angular momentum.
\end{abstract}
\maketitle
\section{Introduction}
Nowadays, the development of protocols of a quantum memory with the ability to store and retrieve recorded quantum states is one of the cornerstones of quantum communications and quantum computing. Many schemes based on the interaction of light with the matter have been proposed. In particular, we can mention schemes based on the EIT \cite{{Lukin2003},{Fleischhauer2005}} effect, the photon echo effect \cite{{Kroll2001},{Kraus2006}}, Raman scattering by $\Lambda$-atoms \cite{Kozhekin2000} and many others. In addition, experimental implementations of devices for high-efficiency storage of the quantum states were developed in \cite{{Appel2008}, {Honda2008},{Chen2013}}.

In recent years, the possibility of not only storing but also shaping a signal in memory cells has been actively discussed. Concerning this necessity, an important requirement for the developed protocols is the efficient writing of a multimode signal.  The possibility to write and shape various time profiles of the quantum field was demonstrated in \cite{{Kuzmin2015},{Manukhova2017}}. With the proper choice of the driving field parameters, the shaping of the signal can be performed with an efficiency close to 1. As for the problem of spatial multimode memory,  we can mention the development of a holographic memory protocol in the resonator configuration for optical images \cite{IVS}, as well as spatial multimode memory in free space \cite{GG2012}. The storage of individual Hermite--Gaussian \cite{Higg2012} and Laguerre--Gaussian \cite{Nicolas2014} modes on ensembles of cold atoms was experimentally demonstrated, but no theoretical analysis of the storage and conversion of such modes on a memory cell was performed.

The Laguerre--Gaussian modes are of particular interest, because, as shown in \cite{Allen92}, they possess a certain orbital angular momentum (OAM). Unlike states with a certain intrinsic momentum (polarization), states with OAM form an infinite-dimensional basis in a Hilbert space. The orbital angular momentum of light can take any integer values, which makes OAM states an indispensable resource for constructing multipartite entangled states.

Existing OAM converters are based on the use of such technical devices as phase holograms \cite{{Heckenberg92},{Karimi2009}}, Dove prisms \cite{Gonzalez06}, cylindrical lens systems \cite{ABRAMOCHKIN1991}. Such schemes, however, consists of a large number of optical elements and their use is accompanied by significant losses. Besides, the effective transformation of modes with different OAM using such optical elements requires to perform some changes of the system parameters specifically for each mode, which may not be satisfactory for many quantum protocols, for example, for quantum computing schemes.

The creation of multipartite quantum states, as well as manipulations with them, often requires mode mixing on linear optical devices such as beamsplitter. However, noiseless mixing of the modes with different spatial profiles on beamsplitters is not possible. In such a procedure, the output fields turn out to be mixed with vacuum noise \cite{Korolev2018}. Within that framework, the conversion of light on a quantum memory cell opens up important possibilities, since it allows linear mixing of different modes without additional vacuum noise.

Given all of the above, we can conclude that the problem of finding new ways to efficiently store and convert quantum states with OAM is relevant in the context of current problems and challenges of quantum optics.

In our work, we are based on the protocol of Raman memory on cold atoms, described in \cite{{Golubeva2011},{Golubeva2012}}. We use the developed approach to the problem of storage and conversion of the Laguerre--Gaussian modes. We will demonstrate the possibility of efficient storage and retrieval of the Laguerre--Gaussian modes by the driving field treated as a plane wave. We will also study the possibility of performing the efficient conversion of the orbital angular momentum of a quantum field on a memory cell and find the optimal configuration of driving fields for such a transformation.

\section{Physical model} \L{2}
In this paper, we consider the Raman model of quantum memory. The possibility of writing quantum statistics of light with OAM on an ensemble of stationary cold atoms is examined. We will study the three-level atoms with $\Lambda $-configuration of energy levels. Atoms interact with a strong classical driving field with the frequency $\omega_d$ and a weak quantum field with certain OAM and frequency $\omega_s$. The frequencies $\omega_d, \; \omega_s$ are detuned from the frequencies of the atomic transitions $\omega_{23} $ and $ \omega_{13}$ respectively by $-\Delta$ . Let us suppose that at the initial moment of time all atoms prepared in the state $|1\> $, and the polarization of the fields $ \vec{\epsilon_d} $ and $ \v\epsilon_s$ are taken so that the driving field acts on the transition $|2\> $-$|3\> $, and quantum one - at the transition $ |1\> $ - $|3\>$ (see Fig. \ref{FIG1}).
\begin{figure}
\includegraphics[scale=0.69]{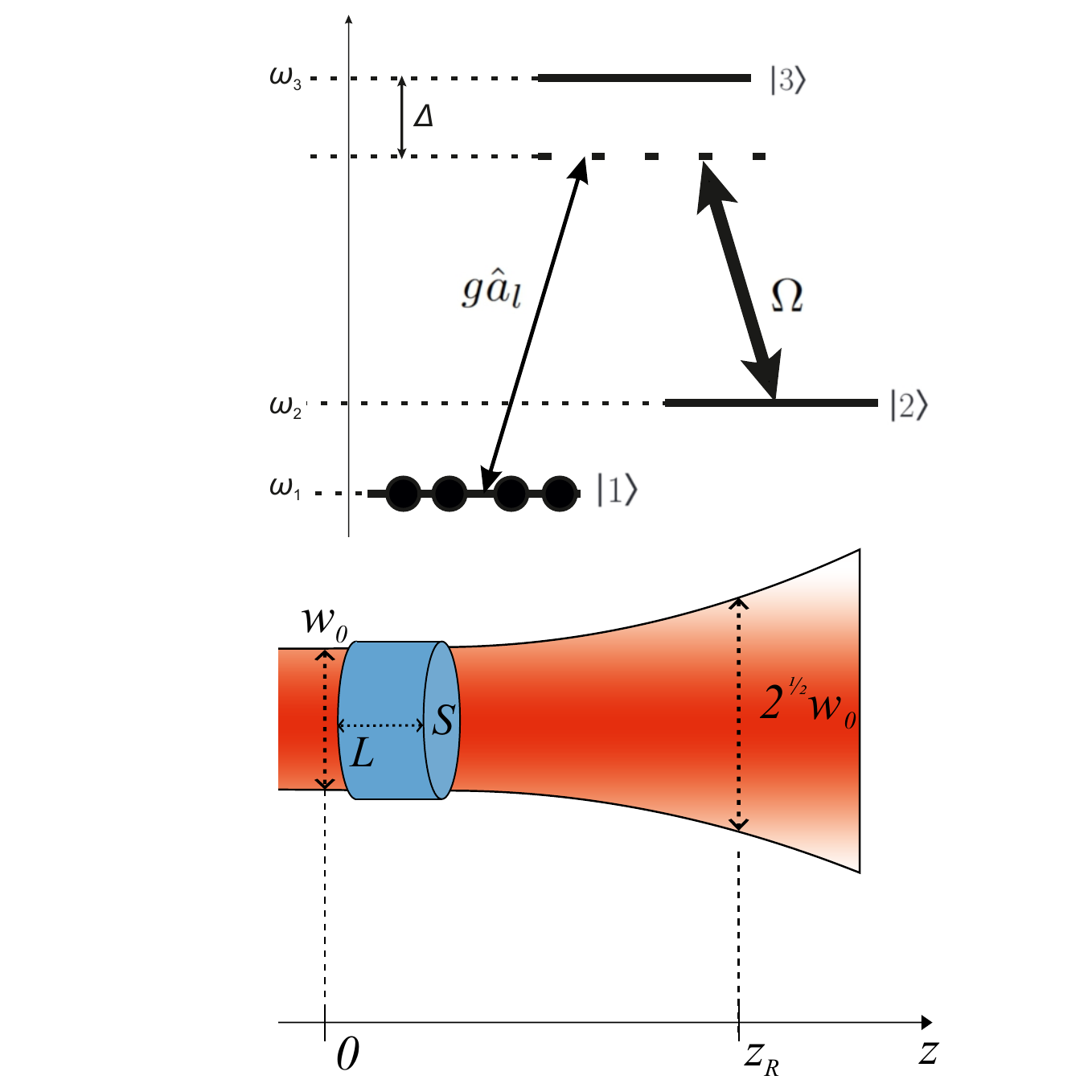}
\caption{A cell with a cloud of cold atoms represented as a cylinder with a length of $L$ and a transverse area of $S$. The cell is located in the plane $z=0$ -- where the wavefront of the wave can be treated as the plane one. The relations between the geometric parameters of the fields and the atomic ensemble is discussed in Section \ref{III}.}\L{FIG1}
\end{figure}

We generally assume that both the driving and quantum fields are quasimonochromatic quasiplane waves propagating along the $ z $ axis. Both waves are considered in the paraxial approximation. Here, we will not yet clarify the spatial dependence of the slowly varying envelope of the driving field $E_0(\v{r}, t) $, and represent the envelope of the quantum field as a set of Laguerre--Gaussian modes:
\BY
\vec{E}_d(\r,t)&=&- i E_0(\r,t)e^{-i \omega_d t + i k_d z}\vec{\epsilon_d} + c.c.,\\
\vec{\hat{E_s}}(\vec{r},t)&=&-i\sqrt{\frac{\hbar\omega_s}{8\pi}}\sum\limits_m{} \;\h a_m(z, t) U^{\perp}_m(\vec{\rho}\;) \nn\\
&&\times e^{-i \omega_s t + i k_s z}\vec{\epsilon_s} + H.c.,
\EY
 where $\rho, \phi, z$ are the cylindrical coordinates, $k_d,k_s$ are the wave numbers corresponding to the carrier wave frequencies, $U^{\perp}_m (\v\rho\;) $ are the Laguerre--Gaussian functions \cite{Allen92} in the plane $z=0$, defined as follows:
\BY
U^\perp_{m}(\v\rho\;)&=&\sqrt{\frac{2}{\pi w^2_0|m|!}}\left(\frac{\rho\sqrt2}{w_0}\right)^{|m|}e^{-\frac{\rho^2}{w^2_0}}e^{im\phi}\L{LG}.
\EY
 Here $w_0$ is the radius of the beam cross section in the plane $z=0$. The functions $U^\perp_{m}(\v\rho)$ form a complete orthonormal set on a plane perpendicular to the direction of the wave propagation:
 \BY
 &&\int d \v{\rho}\; U^{\perp*}_{m}(\v\rho\;)U^\perp_{m^\prime}(\v\rho\;)=\delta_{m,m^\prime},\\
 &&\sum\limits_m U^{\perp*}_{m}(\v\rho^{\;\prime})U^\perp_{m}(\v\rho\;) = \delta^{(2)}(\v{\rho}-\v{\rho}^{\;\prime}).
 \EY
The value $\hbar\omega_s |U^\perp_{m} (\v\rho\;) |^2$ is, in fact, the surface energy density in the mode with the index $m$.

Photon annihilation operators $\h a_m (z, t)$ in the Laguerre--Gaussian mode with index $ m $ are defined so that the average $\<\h a^\dag_m(z, t) \h a_m(z, t) \> $ is the particle number flux in the $U^\perp_{m} (\v \rho \;) $ mode. Such operators obey the following commutation relations:
 \BY
\left[\h a^{}_m(z,t),\h a_{m^\prime}^\dag(z^\prime,t)\right]&=&c\delta_{m, m^\prime}(1-\frac{i}{k_0}\frac{\partial}{\partial z})\delta(z-z^\prime),\;\;\L{comm1}\\
\left[\h a^{}_m(z,t),\h a_{m}^\dag(z,t^\prime)\right]&=&\delta(t-t^\prime).\L{comm2}
\EY

We neglect of diffraction effects when writing the commutation relations (\ref{comm1}). The validity of this neglect will be discussed in section \ref{III}.

A standard approach to quantum memory problems is to describe an atomic ensemble with collective coherence and population operators. Discussing the model of quantum memory, we will follow the work \cite{Golubeva2012}, focusing on the differences associated with the spatial features of the fields. Due to these features,  one should take into account not only the length of the ensemble of atoms along the $ z $ axis when describing an atomic system but also their distribution in the transverse plane (the index $ k $ numbers the atoms):
\BY
\h \sigma_{ij}(\r,t)&=&\sum\limits_{k=1}^{N_{at}}\h \zeta^k_{ij}(t)\delta(\v r-\v r_k),\nonumber\\
\h N_i(\r,t)&=&\h \sigma_{ii}(\r,t)=\sum\limits_{k=1}^{N_{at}}\h \zeta^k_{ii}(t)\delta(\r-\r_k),
\EY
where the operators $\h \zeta_{ij}(t)$ are the projectors of the state $|j\>$ onto the state $|i\> $ at the time $t$: $\h\zeta_{ij}=|i\>\<j|$, $ \omega_{ij}=\omega_i-\omega_j$ is the atomic transition frequency, $\vec{\epsilon_{ij}}$ is the polarization vector of the transition. We have assumed that the vector of the dipole momenta of the transitions $\vec{\epsilon_{13}}$ and $\vec{\epsilon_{23}}$ coincide with the polarization vectors of the signal and driving fields, respectively.
The commutation relations for introduced collective variables can be represented as follows:
\BY
\left[\h \sigma_{ij}(\r,t), \h\sigma_{np}(\r^{\;\prime},t) \right]&=&\left(\delta_{j,n}\h\sigma_{ip}(\r,t)-\delta_{i,p}\h\sigma_{nj}(\r,t)\right)\nn\\
&&\times\delta(\r-\r^{\;\prime}).\L{COM}
\EY
The unperturbed Hamiltonian for the system under consideration can be written in the usual way, while the interaction Hamiltonian has some features related to the spatial structure of the fields. Under the dipole approximation and the rotating wave approximation, the interaction Hamiltonian could be written in the following form:
\BY
&&\h V(\r,t)= \int d\r\; i\hbar g\h\sigma^f_{31}(\r\;)\sum\limits_m\h a^f_m(z,t) U^{\perp}_m(\vec{\rho}\;)\nn\\\L{Ham}
&&-\int d\r\; i\hbar g\h\sigma^f_{13}(\r\;)\sum\limits_m\h a^{f\dag}_m(z,t)U^{\perp*}_m(\vec{\rho}\;)\\
&&+\int d\r\; i\hbar\Omega(\r,t)\{\h\sigma^f_{32}(\r\;) e^{-i\omega_d t+ik_d z}-\h\sigma^f_{23}(\r\;) e^{i\omega_d t-ik_d z}\}.\nn
\EY
Here we introduce the notation for the coupling constant between the atom and the field $ g $ and the Rabi frequency $\Omega$:
$$g =\displaystyle\sqrt{\frac{\omega_s} {8\pi\hbar c}} d_{13}; \; \; \; \Omega(\ r, t)=\displaystyle\frac{E_0 (\r, t) d_{23}} {\hbar}. $$
We also used the notation $d_{ij}$ for the matrix elements of the operator of the dipole momentum of transition between from  level $|i\> $ to $ \<j|$ (for simplicity we will consider these elements to be real numbers).
Since the Heisenberg equations are constructed for fast variables, the Hamiltonian (\ref{Ham}) is written in terms of fast variables (denoted by the superscript $f$) associated with the initial ones as follows:
\BY
&&{\h\sigma_{ij}^f}(\r,t)=\h\sigma_{ij}(\r,t)e^{-i\omega_{ij}t},\\
&&\h a_m^f(z,t) = \h a_m(z,t) e^{ik_sz-i\omega_{s}t}.
\EY

\section{Constraints on the size of the system imposed by neglect of diffraction effects}\L{III}
Since the commutation relations (\ref{comm1}) - (\ref{comm2}) for field operators were derived under the paraxial approximation and without the diffraction effects, we should discuss the conditions under which such approximations are applicable, as well as the restrictions on the size of atomic localization ensembles imposed by these conditions.

We assume that the cell with the atomic ensemble is located in the plane $ z = 0 $, that is, where the wavefront of the wave is flat. Diffraction of a paraxial beam in the near-field zone is described by the two-dimensional Fresnel transform \cite{Murphy}, that is, the field in the $ z $ plane connected with the field in the $ z = 0 $ plane by an integral transformation of the form:
\BY
E(\v{\rho},z)&=&\frac{2\pi ie^{-ikz}}{\lambda z}\int d\v{q}\;\v{q}\;E(q,0)\exp{\{\frac{i k}{2z}(q^2+\rho^2)\}}\nn\\
&&\times J_0(k\v{q}\v{\rho}/z).
\EY
Here $J_0(k\v{q}\v{\rho}/z)$ is the Bessel functions of the first kind.

The Laguerre--Gaussian functions are eigenfunctions of such a transformation, so that we can obtain the explicit form of the function $U_l^{LG}(\v{\rho},z) $ by multiplying the Laguerre--Gaussian function $U_l^\perp(\v{\ rho},0) $ by some eigenvalue:
\BY\L{difro}
U_l^{LG}(\v{\rho},z)&=&U_l^\perp(\v{\rho},0)A(\rho,z)\exp{\{ikz\Phi(\rho,z)\}},\L{difr}\\
A(\rho,z)&=&\left(\frac{\pi^2w_0^4}{\pi^2w_0^4 +\lambda^2z^2}\right)^{\frac{l+1}{2}}\exp{\left\{\frac{\rho^2 z^2\lambda^2}{w_0^2\left(z^2\lambda^2+\pi^2w^4_0\right)}\right\}},\nn\\
\Phi(\rho,z)&=&1+\frac{\rho^2\lambda^2}{2(z^2\lambda^2 + \pi^2 w_0^4)}-\frac{\lambda(|l|+1)}{2\pi z}\arctan{\frac{z\lambda}{\pi w_0^2}}.\nn
\EY
Let assume that a cell with an atomic ensemble is a cylinder with a length of $ L $ and a cross-sectional area of $ S $ (see Fig. \ref{FIG1}). We are interested in the case when the shape of the field's spatial profile does not change at the scale of the cell. From the expression (\ref{difro}) it can be seen that to implement this we need to require the following:
\BY
&&A(\sqrt{S},L)\approx1\L{15},\\
&&\Phi(\sqrt{S},L)\approx1.\L{16}
\EY
We assume that the $L$ is a small parameter: $L\ll\displaystyle\frac{\pi w_0^2} {\lambda} $. In this case, the expressions (\ref{difro}) becomes more convenient:
\BY
A(\v{\rho},L)&=&\exp{\left\{\frac{\rho^2}{w_0^2} \; \left(\frac{L^2 \lambda^2}{\pi^2w_0^4}\right)\right\}},\L{17}\\
\Phi(\v{\rho},L)&=&1+\frac{\rho^2}{2 L^2}\left(\frac{L^2 \lambda^2}{\pi^2w_0^4}\right) -\frac{\lambda^2(|l|+1)}{2\pi^2w_0^2}.\L{18}
\EY

From the expression (\ref{17}) it can be seen that the condition (\ref{15}) is fulfilled if the transverse area of the ensemble $S=\pi\rho^2 $ equal to the transverse area of the beam $\pi w_0^2/4 $. This condition is native for interaction problems since if this condition is not fulfilled, only a part of the beam will interact with the ensemble, which, naturally, will cause losses and decrease the memory efficiency.

In the formula (\ref{18}) we need to choose the parameters in such a way as to be able to neglect the second and third terms. The second term of the phase factor (\ref{18}) disappears if $\rho \sim L $. At the same time, we have already formulated the constraints of the form $ \rho \sim w_0 $. After combining these conditions, we could write down the requirements for the ratio of the beam parameters: $\pi w_0 \gg \lambda $. Since we are considering the optical frequency range, this requirement does not impose significant restrictions and can be considered as satisfied with real experimental parameters. The third term in (\ref{18}) also disappears, taking into account all the approximations described above. As a result, we have the following requirements for the size of the system:
\BY
&L\ll \displaystyle \frac{\pi w_0^2} {\lambda}, \\
&S \approx \pi w^2_0.
\EY
Under these restrictions, we have the right to neglect the diffraction effects and not to take into account changes in the curvature of wave fronts.

\section{Storing and retrieving modes with OAM}\L{IV}

In this section, we consider the procedure for only storing OAM modes without conversion. We want to make sure that the memory protocol for such degrees of freedom could be reduced to the protocols of spatially multimode memory described in the literature \cite{{Golubeva2011},{Golubeva2012}}. Also, our goal is to identify natural variables for describing the interaction of field and atomic systems.
\subsection{Heisenberg equations}
In the framework of this section, we assume that the driving field is a plane monochromatic wave, i.e., the Rabi frequency is independent of spatial and temporal coordinates:
\BY
\vec{E}_d(z,t)&=&- i \frac{\hbar\Omega}{d_{23}} e^{-i \omega_d t + i k_d z}\vec{\epsilon_d} + c.c.
\EY
We use the unperturbed Hamiltonian and the Hamiltonian (\ref{Ham}) to write the Heisenberg equations for field and atomic variables. In this case, we focus on the description of the evolution of each Laguerre--Gaussian field mode. Writing down the equations of interaction between the mode with the number $ l $ and the atomic system, we obtain (further, the arguments of the operators are indicated only where it is necessary to emphasize their presence):
\BY
&&\left(\frac{\partial}{\partial t} + c\frac{\partial}{\partial z}\right)\h a_l=-g\int d\v\rho^{\;\prime}\h\sigma_{13}(\vec{\rho}^{\;\prime},z)U^{\perp*}_l(\vec{\rho}^{\;\prime})\L{H1}\\
&&\dot{\h\sigma}_{13}=-i\Delta\h\sigma_{13}+g  \sum\limits_m \h a_m U^{\perp}_m\left(\h N_1-\h N_3\right)+\Omega\h\sigma_{12}\;\;\;\;\;\;\;\;\\
&&\dot{\h\sigma}_{12}=-g\sum\limits_m\h a_m U^{\perp}_m\h \sigma_{32} - \Omega\h\sigma_{13}\L{SIG}\\
&&\dot{\h\sigma}_{32}=-i\Delta\h\sigma_{32}+g \sum\limits_m\h a^\dag_m U^{\perp*}_m\h\sigma_{12}+\Omega\left(\h N_2-\h N_3\right)\;\;\;\;\;\;\;\;\\
&&\dot{\h N}_{1}=-g \h a_l  U^{\perp}_l\h\sigma_{31} -g\sum\limits_m \h a^\dag_m U^{\perp*}_m\h\sigma_{13} \\
&&\dot{\h N}_{2}=-\Omega \left(\h\sigma_{32}+\h\sigma_{23}\right)\\
&&\dot{\h N}_{3}=-\dot{\h N}_{1}-\dot{\h N}_{2}.
\EY
We will consider the interaction in the Raman limit, assuming that the detuning from the upper excited level $\Delta$ is large, and this energy level is not populated, that is, only two-photon transitions occur. In this case, we are also able not to take into account the relaxation of the population of the third level.

From the equation (\ref{H1}) it can be seen that the evolving variables associated with the mode of the signal field with the number $ l $ are the operators $ \h \sigma^l_{ij}(z, t) $, which could be defined by the expression
\BY
\h\sigma^l_{ij}(z,t)&=&\int d\v{\rho}\;\sigma_{ij}(\vec{\rho},z,t)U^{\perp*}_l(\vec{\rho}).
\EY
However, the commutation relations for these operators have the integral form:
\BY
\left[\h\sigma^l_{ij}(z,t),\h\sigma^{l^\prime}_{ji}(z^\prime,t)\right]&=&\int d\v{\rho}\;\left( \h N_i(\vec{\rho},z)-\h N_j(\vec{\rho},z)\right)\nn\\
&&\times U^{\perp*}_l(\v\rho\;)U^{\perp}_{l^\prime}(\v\rho\;)\delta(z^\prime-z^\prime).\L{commut}\;\;\;\;\;
\EY

It follows from the commutation relations that in the general case such variables cannot be considered as independent from each other, which is extremely inconvenient for describing the system. However, as will be seen later, this nuisance can be avoided using standard approximations of quantum memory problems.

Let's recall that the atoms in the ensemble initially were predominantly in the state $|1\>$. Considering that the quantum field is rather weak, we can say that the number of atoms in the state $ |1\> $ does not change significantly during the interaction and the population of the level $ |1\> $ remains much larger than the population of the level $|2\> $. This leads to the condition $\bar{\h{\sigma}}_ {32} \ll \bar{\h{\sigma}}_ {13} $, which allows us to neglect  the first term relative to the second in the  equation (\ref{SIG}).

The significance of the population of the level $|1\> $ compared with the populations of other levels, also gives us the opportunity to say that $\bar{\h N}_1- \bar{\h N}_3 \approx \bar {\h N}_1 \equiv N $, where $ N $ is the average concentration of atoms evenly distributed in the cell. Then we can use the Laguerre--Gaussian mode orthogonality and remove the integral dependence in (\ref{commut}). The commutation relations for the coherence operators of interest $ \h \sigma^{l}_{13}, \h \sigma^{l}_{12} $ can be rewritten in the form:

\BY
\left[\h\sigma^l_{ij}(z,t),\h\sigma^{l^\prime}_{ji}(z^\prime,t)\right]&=&N\delta_{l,l^\prime}\delta(z-z^\prime),\\
\{i,j\}&=&\{1,2\};\{1,3\}.\nn\;\;\;
\EY

All written above allow us to close the system of equations for operators $\h a_l, \h \sigma_{13}^l$ and $\sigma_{12}^l$:
\BY
&&\frac{\partial}{\partial z}\h a_l=-g\sqrt{N} \h c_l\L{HF1}\\
&&\dot{\h c}_l=-i\Delta\h c_l+g \sqrt{N} \h a_l+\Omega\h b_l\;\;\;\;\;\;\;\;\L{HF2}\\
&&\dot{\h b}_l= - \Omega\h c_l.\L{HF3}
\EY
Here the normalized operators $\h c_l$ and $\h b_l$ are introduced as following:
\BY
&&\h c_l(z,t) = \h\sigma^l_{13}(z,t)/\sqrt{N};\;\;\;\;\h b_l(z,t) = \h\sigma^l_{12}(z,t)/\sqrt{N}\;\;\;\;\;\;\;.
\EY
 We assume that the pulse duration $ T $ is large enough to neglect the time of the propagating of the pulse wave fronts ($ cT \gg L $) through the medium. Then we can formally consider the time derivative in the equation (\ref{H1}) to be small in comparison with the other terms.

It should be mentioned that in the resulting system, all spatial modes of spin coherence with different OAMs evolve independently of each other.

The equations (\ref{HF1}) - (\ref{HF3}) coincide with the equations usually used to describe the quantum memory \cite{Golubeva2012}. An important feature of our consideration is that modes with different OAM interact with different spatial modes of spin coherence. When considering the spatially homogeneous distribution of atoms, the bosonic operators $ \h b_l (z, t), \h c_l (z, t) $ can be considered as the annihilation operators in spin coherence modes with a certain OAM.
\subsection{Solutions for writing and readout processes}\L{IVB}
For completeness, we write out the solutions of the system (\ref{HF1})--(\ref{HF3}). To do this, we pass from the operator quantities to the c-number ones and omit all the vacuum terms, bearing in mind that we are interested only in normally ordered means. For convenience, we will also use dimensionless variables:$$\tilde t= \Omega t;\;\;\;\tilde z = \displaystyle\frac{2g^2 N}{\Omega}z.$$.

We can obtain the following expression for the spin coherence at the writing stage:
\BY
b^W_l(\tilde{z}, \tilde{t}) &=& -\frac{g \sqrt{N}}{\Omega}\int\limits_0^{\td{T}_W} d\tilde{t}^\prime a_l(0,\tilde{t}^\prime) G_{ba}(\td{z},\tilde{t}-\tilde{t}^\prime).\L{WR}
\EY
Here $ \tilde{T}_W $ is the effective duration of the writing pulse, and $G_{ba}$ is the kernel of the integral transformation \cite{Golubeva2012}. We assume that during storage the coherence between the levels $  1> $ and $ |2> $ is preserved without any losses ($ b^R_l (\tilde {z}, 0) = b^W_l (\ tilde{z}, \td {T}_W $), and the coherence between the levels $|1> $ and $|3> $ decays, so at the beginning of readout process $c_l^R (\tilde{z}, 0) = 0 $. Then for the readout stage we can write the following expression:
\BY
&&a^R_l(\tilde{L}, \tilde{t}) = -\frac{\Omega}{2g \sqrt{N}}\int\limits_0^{\td{L}} d\tilde{z}\; b^W_l(\tilde{z}, \tilde{t}) G_{ab}(L-\tilde{z}, \td{t})\;\;\;\nn\\
&&=\int\limits_0^{\td{T}_W} d\tilde{t}^\prime a_l(0,\tilde{t}^\prime)K(\td{t},\td{t}^{\prime}),\L{RD}\\
&&K(\td{t},\td{t}^{\prime})=\int\limits_0^{\td{L}} d\tilde{z}\;G_{ba}(\td{z},\tilde{t}-\tilde{t}^\prime) G_{ab}(\td{L}-\tilde{z}, \td{t}).
\EY
Here $K (\tilde {t}, \tilde {t}^\prime)$ is the kernel of the full memory cycle (in the configuration of the copropagating fields during writing and readout processes). Note that the kernel $K (\tilde {t}, \tilde{t}^\prime)$ does not depend on the index $ l $ and completely coincides with the kernel described in \cite{Golubeva2012}. For this reason, we do not provide here an analysis of the efficiency of the storage protocol, which was done in the cited work.

 An important result is the following observation: if the driving field is a plane wave, we can write a quantum field with a specific OAM on the atomic medium and subsequently retrieve the field without changing the spatial profile. In the process of interaction between the driving and quantum fields with the atomic medium, the quantum - statistical properties of each mode with OAM are "written" \; \; to the corresponding spatial mode of spin coherence. Hence, we are able to say that in the process under consideration, the field modes $ \h a_l $ with different projections of the orbital angular momentum $ l $ evolve independently of each other.
\section{Converting modes with OAM}
In this section, we analyze the possibility of not only storing, but simultaneously converting quantum states with OAM in the quantum memory protocol. We will also identify the optimal values of the driving field parameters for the optimum conversion.

\subsection{Heisenberg equations of the system with the complex spatial structure of the driving field}
Let us now consider the case when the driving fields at the writing and readout stages are Laguerre--Gaussian modes in the plane $ z = z_ {S} $ with OAM projections equal to $ J $ and $ I $, respectively:
\BY
&&\vec{\h {E}}^W_d=i B_J U_J^\perp(\v\rho\;,z_{S})e^{ik_dz-i\omega_dt}\vec{\epsilon_d} + c.c.\L{39}\\
&&\vec{\h {E}}^R_d=i B_I U_I^\perp(\v\rho\;,z_{S})e^{ik_dz-i\omega_dt}\vec{\epsilon_d} + c.c.\L{40}
\EY

The amplitudes of the driving fields will be considered to be real numbers for simplicity. The spatial profile of the Laguerre--Gaussian mode with the momentum $ l $ in the plane $ z = 0 $ is a ring of radius $w_0\sqrt{(| l | +1)}/2 $ \cite{Phillips83} with zero field intensity in the center of the transverse plane. Since that overlapping of the modes with different momenta is rather weak. At the same time, the presence of both the driving and quantum fields in every point of an ensemble is required to ensure two-photon transitions. To do this, we can consider beams of different radius $ w_0 $, however, for clarity, we instead introduce the parameter $z_ {S}$ - the distance between the waists of the signal and driving beams (see Fig. \ref{ZEFF}). This parameter allows us to follow the overlapping of fields. By varying the parameter $ z_S $, we essentially change the ratio of the radii of the beams of the driving and signal fields in the plane $ z = 0 $, and thereby try to ensure the best overlap between the modes.

\begin{figure}
\includegraphics[scale=0.38]{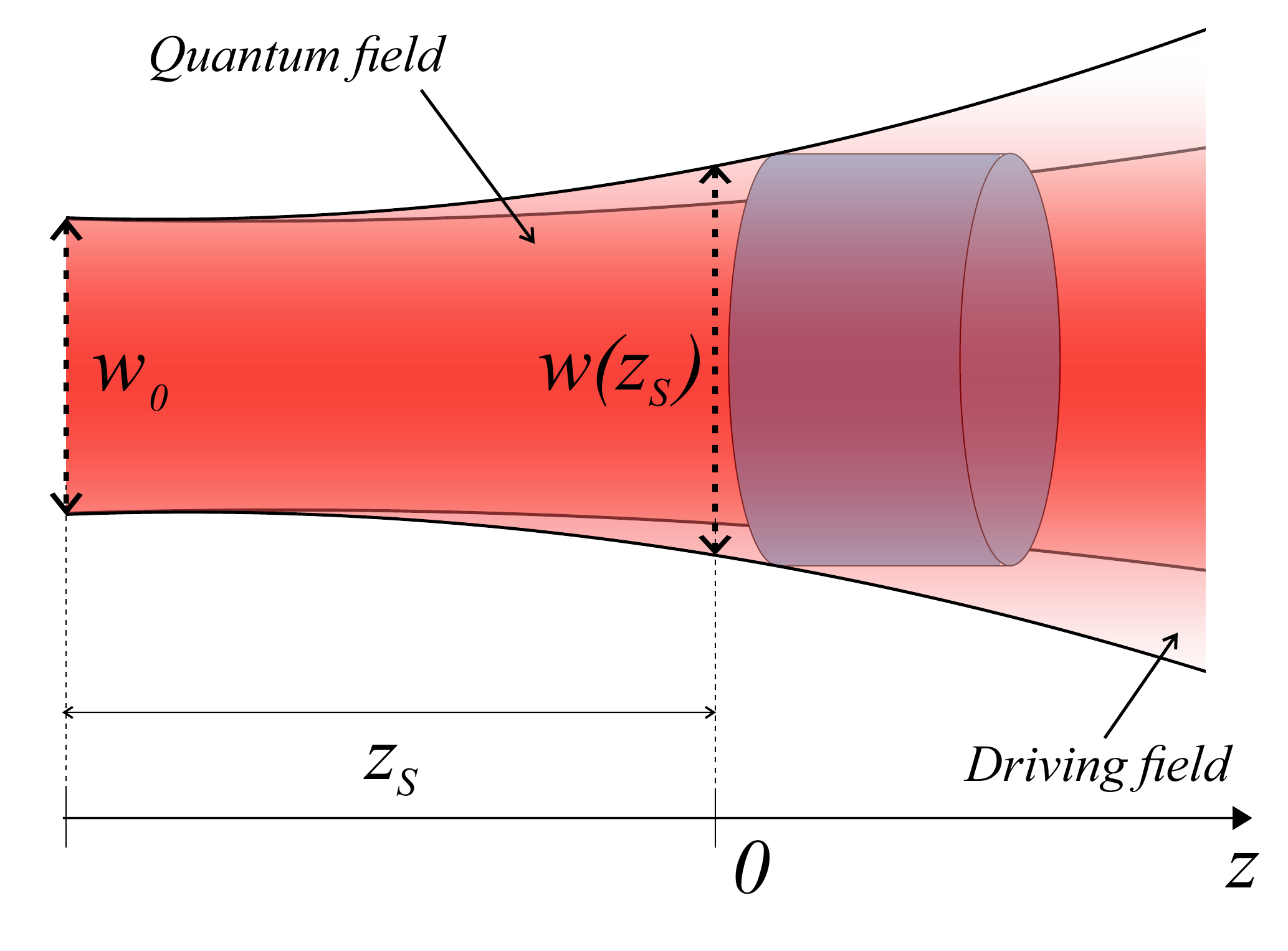}
\caption{To ensure good mode overlap, we consider a field geometry such that the waist of the driving beam and the waist of the signal beam are shifted one from another on the value
$z_S$.}\L{ZEFF}
\end{figure}

 Let's recall also that the operator and classical field amplitudes introduced in the section \ref{2} were normalized so that their square had the meaning of a photon flux per unit time through a plane perpendicular to the propagation direction.  If we want to discuss the interaction the classical and quantum modes with different transverse sizes, we should operate the equally normalized quantities. In the expressions (\ref{39}) and (\ref{40}), we divide the amplitudes $B_J $ and $B_I $ by the area of the Laguerre--Gaussian beams with the corresponding momentum: $ B_J\rightarrow~B_J/S_J, \; B_I \rightarrow B_I / S_I $. The squares of the moduli of such new amplitudes are of the energy flux through the unit area, and, thus, are ultimately normalized equally regardless of the mode number. To preserve field dimensionality it is necessary to multiply the functions $ U_J^\perp(\v{\rho}, z_S) $ and $ U_I^\perp (\v{\rho}, z_S) $ by the same factors: $U_J^\perp(\v{\rho}, z_S) \rightarrow~U_J^\perp(\v{\rho}, z_S) S_J, \; \; U_I^\perp (\v{\rho}, z_S) \rightarrow~U_I^\perp (\v {\rho}, z_S) S_I $. Thus we write down fields through normalized operators and dimensionless mode functions. We will do the same with the quantum field. Calculating the area of the Laguerre--Gaussian beam with the azimuthal number $ l $ according to the formula $\displaystyle S_l = \pi w^2_0 (1 + z^2 /z_R^ 2)\frac{(l + 1)}{4} $ \cite{Phillips83}, we re-designate field operators and functions as follows (we denote old and new variables in the same way for compact notation, but hereinafter we operate renormalized quantities):
\BY
&&\frac{\h{a}_l}{\sqrt{\pi w^2_0\frac{(l+1)}{4}}}\rightarrow\h{a}_l,\;\; U^\perp_l(\v{\rho})\sqrt{\pi w^2_0\frac{(l+1)}{4}}\rightarrow{U}^\perp_l(\v{\rho}),\;\;\;\;\;\;\;\L{41}\\
&&\frac{{B}_m}{\sqrt{\pi (1+z_{S}^2/z_R^2) w^2_0\frac{(m+1)}{4}}}\rightarrow{B}_m,\\
&&U^\perp_m(z_{S})\sqrt{\pi (1+z_{S}^2/z_R^2) w^2_0\frac{(m+1)}{4}}\rightarrow {U}^\perp_m(z_{S})\L{43},
\EY
The  orthonormality of the new Laguerre - Gaussian functions and new commutation relations for field operators could be written as follows:
\BY
&&\int d\v{\rho\;} U^{\perp*}_m(\v{\rho}\;,z) U_m^\perp(\v{\rho}\;,z)=\pi w^2_0(1+z^2/z_R^2)\frac{(m+1)}{4}\nn\\
&&=S_m,\;\;\;\;\;\;\;\\
&&\left[\h a^{}_l(z,t),\h a_{l^\prime}^\dag(z^\prime,t)\right]=\frac{1}{S_l}\delta_{l, l^\prime}(1-\frac{i}{k_0}\frac{\partial}{\partial z})\delta(z-z^\prime).\;\;\;\;\;\;\;\;\;\L{comm1N}
\EY

Using the approximations and notation introdused in the previous section we can write the Heisenberg equations for atomic variables and a field variable taking into account the spatial structure of the control field:
\BY
&&\frac{\partial}{\partial z}\h a_l=-\frac{g}{S_l} \int d\v{\rho}^{\;\prime}\sigma_{13}(\v{\rho}^{\;\prime},z,t)U^{\perp*}_l(\vec{\rho}^{\;\prime})\L{46}\\
&&\dot{\h\sigma}_{13}=-i\Delta\h\sigma_{13}+g {N} \sum\limits_p\h a_pU^\perp_p+\Omega_mU_m^\perp(z_{S})\h\sigma_{12}\;\;\;\;\;\;\;\;\\
&&\dot{\h\sigma}_{12}= - \Omega^{*}_mU_m^{\perp*}(z_{S})\h\sigma_{13}.\L{48}
\EY
Here $m = J$ during writing process, and $m = I$ during readout process,  Rabi frequency is denoted as $\Omega_m=B_md_{23}\hbar^{-1}$. In the \ref{IV} section,  the collective operators $\h b_l $ and $\h c_l$ were the natural variables for the system. Therefore, we move on to these variables, decomposing all atomic variables to the set of functions $ U^\perp_l(\v{\rho})$, and try to write down a closed system of equations. Taking into account the new commutation relations of field operators and the normalization of the functions $ U^\perp_l(\v{\rho}\;) $, we rewrite the system of equations (\ref{46})--(\ref{48}):
\BY
&&\frac{\partial}{\partial z}\h a_l=-g \sqrt{N}\h c_{l}\L{HH1}\\
&&S_l\dot{\h c}_{l}=-i\Delta S_l\h c_{l}+g\sqrt{N}S_l\h a_l\nn\\
&&+\Omega_m\sum\limits_n\h b_n\int d\v{\rho}\; U_n^\perp(\v\rho\;)U_m^\perp(\v\rho\;,z_{S})U_l^{\perp*}(\v\rho\;)\;\;\;\;\L{HH2}\\
&&S_{l}\dot{\h b}_{l}=-\Omega_m\sum\limits_n \h c_{n}\nn\\
&&\times\int d \v{\rho}\;U_n^\perp(\v\rho\;)U_m^{\perp*}(\v\rho\;,z_{S})U_l^{\perp*}(\v\rho\;).\;\;\;\;\;\;\;\;\L{HH3}
\EY

Comparing the obtained equations with the equations (\ref{HF1})--(\ref{HF3}), we see that the written system is not closed, since in the equations (\ref{HH2}) and (\ref{HH3}) there are terms containing not only  the operators $ \h b_l $ and $ \h c_l $, but also the projections of spin coherences on all other Laguerre--Gaussian modes. Let's look back on the definition of the functions $U_l^\perp$(\ref{difr}) and write the overlap integrals $\int d\v{\rho}\; U_m^\perp U_n^\perp U_l^{\perp*} $ in explicit form:
\BY
&&\int d\v{\rho}\; U_n^\perp(\v\rho\;)U_m^\perp(\v\rho,\;z_{S})U_l^{\perp*}(\v\rho\;)=\int\rho\; d{\rho}\;d\phi\;\nn\\ &&\times|U_n^\perp(\rho)||U_m^\perp(\rho, z_{S})||U_l^{\perp}(\rho)|e^{i\Phi(\rho, z_{S})}\exp{\{i\phi(n+m-l)\}}\nn\\
&&=2\pi\int\rho\; d{\rho}\;|U_n^\perp(\rho)||U_m^\perp(\rho,z_{S})||U_l^{\perp}(\rho)|e^{i\Phi(\rho, z_{S})}\delta_{n,l-m}\nn\\
&&=\chi_{l,m}\delta_{n,l-m}.\L{CHIL}
\EY
The coefficients $\chi_{l, m} $ can be, generally speaking, complex numbers due to the presence of the phase factor $\exp{\{i\Phi(\rho, z_{S})\}} $ in the function $U_m^\perp (\rho, z_{S}) $ (for the explicit form look at (\ref{difr})), which is define  the spatial profile of the driving field. However, as will be shown later, the complexity of these coefficients will not fundamentally affect the quality of storage and conversion of the quantum field.

The Kronecker delta in (\ref{CHIL}) allows us to remove the sum over $n$ in (\ref{HH2}) and (\ref{HH3}) and write down the following expressions:
\BY
&&\sqrt{S_l}\frac{\partial}{\partial z}\h a_l=-g\sqrt{N}\sqrt{S_l}\;\h c_l\L{HHF1}\\
&&\sqrt{S_l}\;\dot{\h c}_l=-i\Delta\sqrt{S_l}\;\h c_l+g \sqrt{N}\sqrt{S_l}\;\h a_l\nn\\
&&+\Omega_m\frac{\chi_{l,m}}{\sqrt{S_l}\sqrt{S_{l-m}}}\h b_{l-m}\sqrt{S_{l-m}}\;\;\;\;\;\;\;\;\L{HHF2}\\
&&\sqrt{S_{l-m}}\dot{\h b}_{l-m}= - \Omega_m\frac{\chi^*_{l,m}}{\sqrt{S_l}\sqrt{S_{l-m}}}\h c_{l}\sqrt{S_l}.\L{HHF3}
\EY
One can notice that the system of equations is closed now and describes the interaction of the field mode with the number $l$ and the coherence modes with the numbers $l$ and $l-m$, where $m$ is the angular momentum of the driving field (we assume $m=J$ when writing process is discussed and $m=I$ during readout process). Moreover, since the effective Rabi frequency $\Omega_{m}\chi_{l, m}$ depends on the Laguerre--Gaussian overlap integrals with indices $ l, m $ and $ l-m $, the interaction occurs with different efficiencies for different OAM projections of the driving and quantum fields.

\subsection{Driving fields with different OAM for signal field conversion}
To compare the obtained results with the results of section \ref{IV}, we choose the amplitudes of control fields $ B_m $ so that $ \Omega_m = \Omega $. Then the expression in dimensionless variables for spin coherence at the end of the writing process is written as follows:
\BY
b^{W}_{l-J}(\tilde{z}, \tilde{t})&&=-\displaystyle\frac{g \sqrt{N}}{\Omega}\frac{\chi_{l,J}}{S_{l-J}}\int\limits_0^{\td{T}_W} d\tilde{t}^\prime a_l(0,\tilde{t}^\prime)\nn\\
&&\times G^{lJ}_{ba}(\td{z},\tilde{t}-\tilde{t}^\prime).\L{56}\;\;\;\;\;\;
\EY
The dimensionless variables are introduced in the same way as in the section \ref {IV}.
The kernel $ G_ {ba}^{lJ} $ is now characterized by upper and lower indices and depends on normalized overlap integrals (see Appendix \ref {B}). We will perform the analysis of transformation kernels below.

From the expression (\ref {56}) we can conclude that if the driving field has a certain orbital angular momentum of $m$, then the quantum field modes $\h a_l $ do interact not with the corresponding spin coherence modes $\h b_l $ as it was in section \ref{IVB}, but with modes with "shifted" \; index $\h b_{l-m} $. A qualitative interpretation of this result may consist in the fact that the driving field, interacting with the atomic medium, gives us some "reference frame" \; \; -- the phase portrait relative to which the quantum field is considered. Thus, the phase profile of a quantum field is already considered from the "rotating" \; with the speed $m$  reference frame, whence the difference phase coefficient $ l-m $ arises.

We can write solutions for the readout stage for  the same configuration of the fields:
\BY
&&a^{R}_{l+I-J}(\tilde{L}, \tilde{t})= -\frac{\Omega}{g \sqrt{N}}\frac{\chi_{l+I-J,I}}{S_{l+I-J}}\int\limits_0^{\td{L}} d\tilde{z}\; b^W_{l-J}(\tilde{z}, \tilde{t})\nn\\ &&\times G^{lI}_{ab}(\td{L}-\tilde{z}, \td{t})=
C_{IJ}^l\int\limits_0^{\td{T}_W} d\tilde{t}^\prime a_l(0,\tilde{t}^\prime)K^{IJ}_{l}(\td{t},\td{t}^{\prime}),\;\;\;\;\;\L{57}\\
&&C_{IJ}^l=\frac{\chi_{l-J,I}}{S_{l+I-J}}\frac{\chi_{l,J}}{S_{l-J}},\\
&&K^{IJ}_{l}(\td{t},\td{t}^{\prime})=\int\limits_0^{\td{L}} d\tilde{z}\;G^{lI}_{ba}(\td{z},\tilde{t}-\tilde{t}^\prime) G^{lJ}_{ab}(\td{L}-\tilde{z}, \td{t}).
\EY

An important result is that the orbital angular momentum of the retrieved field differs from that of the stored field by the value $ I-J $, where $ J $ is the OAM of the writing driving field and $ I $ is the OAM of the readout driving field. Thus, we have shown that it is possible to perform the transformations of OAM of a quantum field on a quantum memory cell.

The kernels of integral transforms depend on the squares of the normalized overlap integrals as follows:
\BY
&&G^{lm}_{ba}(\tilde{z},\td{t})=G^{lm}_{ab}(\tilde{z},\td{t})=[f^{lm}_0(\tilde{z},\td{t},r)*f^{lm*}_0(\tilde{z},\td{t},-r)]\L{60}\;\;\;\;\;\\
&&f^{lm}_0(\tilde{z},\td{t})=\exp{\{-i\left(\sqrt{r^2+|\chi_{l,m}|^2/S^2_{l-m}} + r\right)\td{t}\}}\nn\\
&&\times J_0\left[\sqrt{\td{z}\td{t}\left(1+\frac{r}{\sqrt{r^2+|\chi_{l,m}|^2/S^2_{l-m}}}\right)}\right]\Theta(\td{t}),\L{61}\\
&&m=J,I.\nn
\EY
Here we use the notation for the dimensionless detuning $ r = \displaystyle\frac{\Delta }{2 \Omega} $. It can be seen from these expressions that even with the complex coefficient $\chi_{l, m} $, the arguments of the Bessel function of the first kind $ J_0 $ remain real. This is an important property of the transformation under consideration since if the transformation kernels (\ref{57}) were complex, this would lead to mixing of the quadratures of the quantum field, which is destructive when considering the issues of storing the quadrature-squeezed states. But, since the kernel $ K^{IJ}_l $ is real, this mixing does not occur in the problem under consideration.

The formulas (\ref{60}), (\ref{61}) greatly simplify in the Raman limit. So, according to the expressions (\ref{F1}), (\ref{F4}), the kernels of the integral transforms $ G_{ba}^{lm}$ and $G_{ab}^{lm} $ in this case completely coincide with kernels $ G_{ba} $ and $ G_{ab} $ from the \ref{IVB} section:
 \BY
 &&G_{ba}^{lm}\xrightarrow[\Delta\gg\Omega]{}G_{ba},\;\;
G_{ab}^{lm}\xrightarrow[\Delta\gg\Omega]{}G_{ab}.
 \EY
Then only the coefficients $C_{IJ}^l $ depend on the OAM of the modes participating in the process:
\BY
 &&a_{l+I-J}(\td{L},\td{t})=C_{IJ}^l\int\limits_0^{\td{T}_W} d\tilde{t}^\prime a_l(0,\tilde{t}^\prime)\nn\\
 &&\times\int\limits_0^{\td{L}} d\tilde{z}\;G_{ba}(\td{z},\tilde{t}-\tilde{t}^\prime) G_{ab}(\td{L}-\tilde{z}, \td{t}).\L{64} \;\;
\EY

The quality of the conversion depends on the eigenvalues of the memory conversion kernel and the coefficients $C_ {IJ}^l $. Kernel analysis has already been done in \cite{Golubeva2012}, but our goal is to study the differences associated with the presence of a coefficient in the integral (\ref{64}).

It follows from the form of (\ref{64}) that each of the stages (writing and readout) degrades the signal quality $\chi_{l, m}/S_{lm} $ times (since the normalized overlap integrals are smaller than 1). Let's change the configuration of the fields in such a way as to carry out the conversion only at one stage (either during writing or during readout). At another stage then we should perform the writing/readout of the signal without conversion. This will make the coefficient before the integral in (\ref{64}) linear in $\chi_{l, m}/S_{l-m} $. From section \ref{IV} we know that storing without conversion can be done using a plane wave. Therefore, if one firstly performs a conversion by a driving field with OAM, and then retrieve it with a plane wave, the quality of the conversion is greatly improves.

\subsection{Efficiency of the conversion of states with different OAM}
Let's continue the discussion of the previous section, we consider the situation when the writing field $ \vec{\h{E}}^W_d (\v{r}, t) $ has a certain OAM, and the readout one  $\vec {\h{E}}^R_d(\vec{r}, t) $ is a plane wave:
\BY
\vec{{E}}^W_d(\v{r},t)&=&-i B_J U_J^\perp(\v\rho\;,z_{S})e^{ik_dz-i\omega_dt}\vec{\epsilon_d} + c.c.,\;\;\;\;\;\;\;\;\;\\
\vec{{E}}^R_d(\vec{r},t)&=&-i E_0e^{-i \omega_d t + i k_d z}\vec{\epsilon_d} + c.c.,\\
\vec{\hat{E_s}}(\vec{r},t)&=&-i\sqrt{\frac{\hbar\omega_s}{8\pi c}} \;\sum\limits_p\h a_p(z, t) U^{\perp}_p(\vec{\rho}\;)\nn\\
&&\times e^{-i \omega_s t + i k_s z}\vec{\epsilon_s} + H.c.
\EY
Here, we still use dimensionless Laguerre--Gaussian functions and normalized amplitudes $B_J $ and $ \h a_l $ introduced in the previous section (\ref{41} - \ref{43}). We set amplitudes of the driving fields equal to each other and consider them as a real numbers: $ E_0 = B_J = E^*_0 = B^{*}_J $. Omitting the reasoning similar to this presented in previous sections, we write down the relation connecting the field retrieved after the readout stage with the original signal:
\BY
a_{l-J}(\td{L},\td{t})&=&\frac{\chi_{l,J}}{S_{l-J}}\int\limits_0^{\td{T}_W} d\tilde{t}^\prime a_l(0,\tilde{t}^\prime) K(\td{t},\td{t}^{\prime}).\;\;\L{68}
\EY

As one can see, with such a configuration of driving fields, we could perform the transformation of the orbital angular momentum of the quantum field. So, if one writing a field with a certain momentum $ l $ on the memory cell,  then the retrieved field's OAM is considered to be $ l-J $. Similarly, if writing is carried out by a plane wave, and readout process by a field with OAM, then the orbital momentum of the retrieved field will be $ l + J $.

The optimization of the considered memory protocol for spatial modes by selecting the effective cell length and temporal field profiles was considered in \cite{Golubeva2012}. Further, we will assume that the values of these parameters correspond to the best process efficiency, i.e. the integral transformation with the kernel $K (\td{t}, \td{t}^{\prime}) $ can be replaced by its eigenvalue equal to one. We focus here on a detailed analysis of "quality" \;  of the retrieved Laguerre -- Gaussian modes. This quality depends on the OAM of the signal and driving fields, that is, on the overlap integrals $\chi_{l, J}/S_{l-J}$. Let's recall, that we shifted the driving and quantum field waists by a certain amount $ z_ {S} $ to ensure good mode overlap, considering all other parameters (the radius $ w_0 $ and the Rayleigh range $ z_R $) of the beams to be the same.

We first consider the case of storing without conversion. We choose a driving field at the writing stage in the form of a Gaussian beam with OAM $ J = 0 $. As can be seen from (\ref{68}), this field does not change the index $ l $ of the retrieved signal, however, the process efficiency (in contrast to the case of writing and readout by a plane wave) is determined by the corresponding coefficients $ \chi_{l, 0}/S_0 $. Let us analyse the dependence of these coefficients on the relative shift of the beams $ z_S / z_R $ (see Fig. \ref{FIGCHI}).

 \begin{figure}[h!]
\includegraphics[scale=0.4]{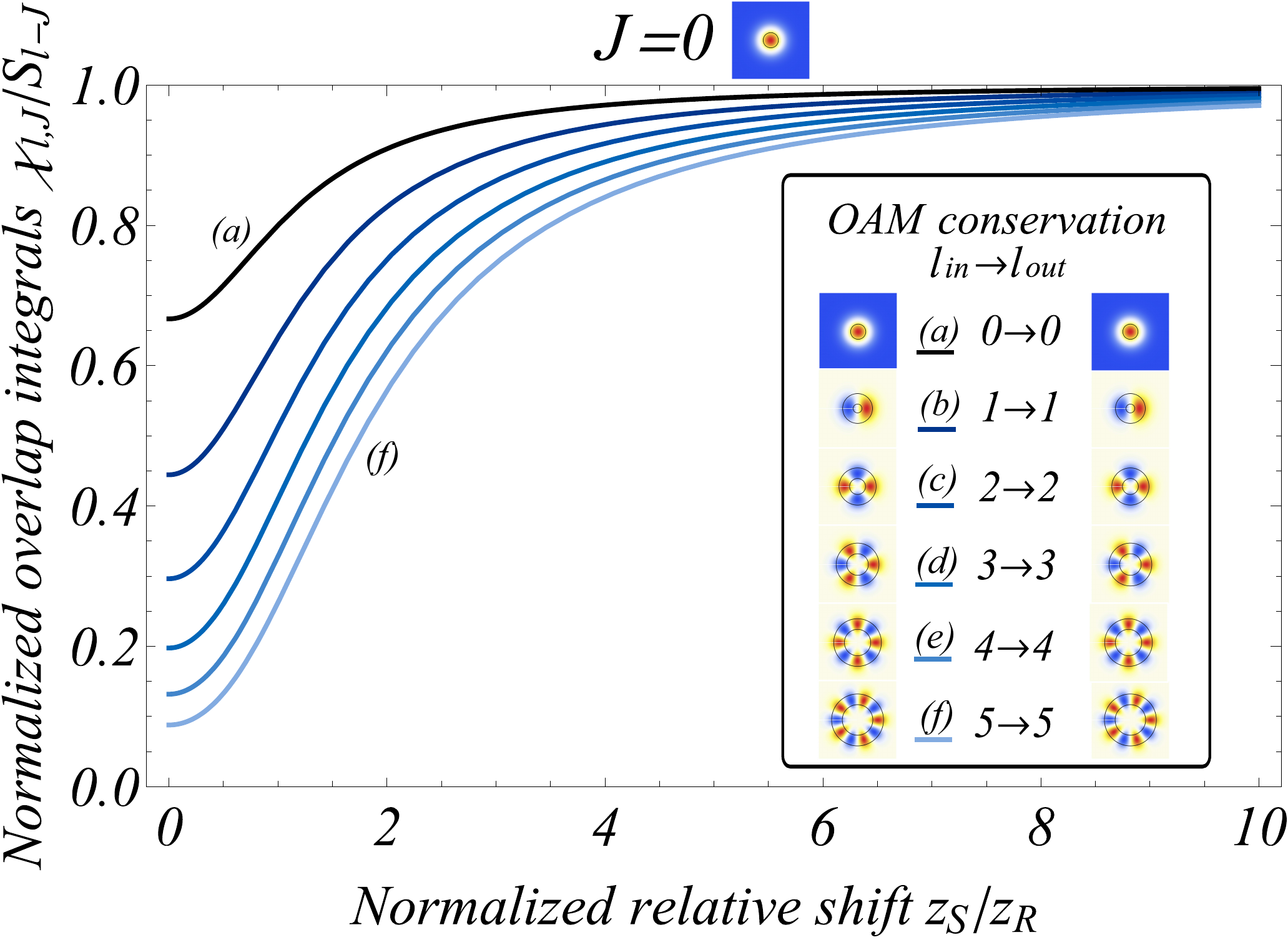}
\caption{The dependence of the normalized overlap integrals on the OAM of the signal field for the case of storage without conversion $ J = 0 $. Values close to 1 are achieved when the beam area of the driving field is much larger than that of the signal field.}
\L{FIGCHI}
\end{figure}

As expected, in the limit $z_{S}\gg z_R $ the conversion coefficient tends to unity, as if we were performing writing and readout by a plane wave, since with extreme values of the parameters the spatial inhomogeneity of the driving Gaussian field is vanishing, $ \displaystyle \exp{\{-\frac {\rho^2} {w^2 (z_S)} \}} \rightarrow1 $. An interesting fact is that the coefficient $\chi_{0,0} / S_ {0} $ is not equal to one when $z_ {S} = 0 $, $l = J = 0$, that is, when we consider the interaction of two Gaussian beams of the same transverse area. This result can be explained by the fact that the amplitude of the driving field, as well as the amplitude of the quantum field, decreases with distance from the point $\rho = 0$ in the same way as the Gaussian function. In this case, the mode of spin coherence has an effective transverse size $ \sqrt{2} $ times smaller than the initial transverse field size, since its amplitude in the cross-section will decrease two times faster than the amplitudes of the initial fields due to the overlap of two Gaussian profiles. Then the mode of the retrieved quantum field will also have a smaller effective transverse size compared to the original signal. Summarizing the above, it can be argued that the interaction of fields with different spatial profiles changes the mode composition of the signal so that the basis set of the field at the output of memory cell will be different from that of the field at the input. However, as we show below, this difference can be small. Therefore, the problem of the conversion of one mode to another in the same basis set remaining valid.

 Considering the question of converting the OAM of a quantum field, we first analyze the possibility of retrieving light with the OAM from a memory cell when at the input of the cell we have a quantum field  that does not have OAM, i.e. with $ l = 0 $ and a driving field with momentum of $ J, J = 1,2,3, ... $.  Fig. {\ref{FIGCHI1}} shows the normalized overlap integrals  against the ratio $ \displaystyle\frac{z_{S}}{z_R} $ for different values of the OAM of the driving field.
 \begin{figure}[h!]
\includegraphics[scale=0.4]{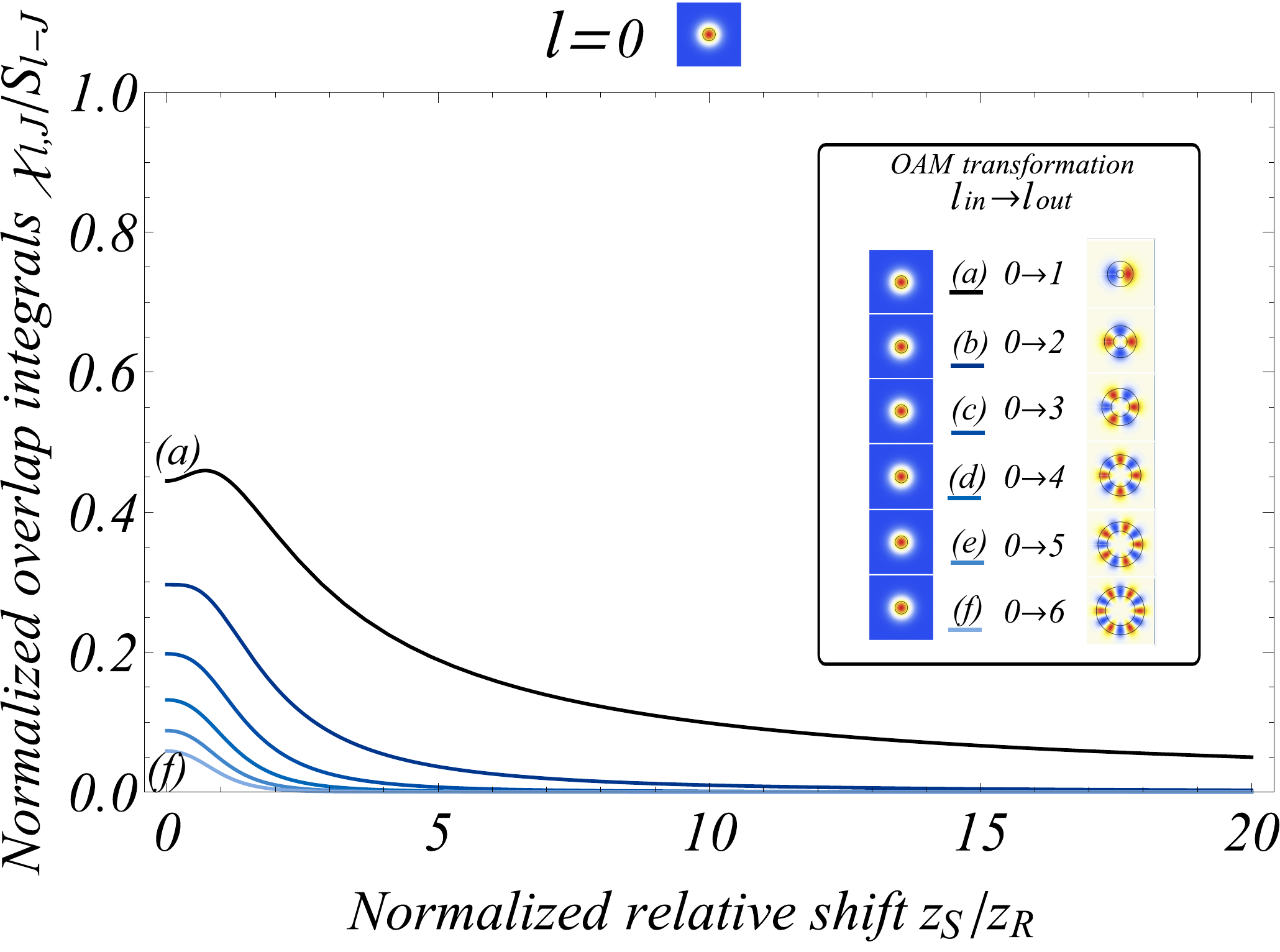}
\caption{Normalized overlap integrals depending on the normalized relative shift for different values of the momentum of the driving field from $ J = 1 $ (curve (a)) to $ J = 6 $ (curve (f)).}
\L{FIGCHI1}
\end{figure}

From Fig. {\ref {FIGCHI1}} we can notice that the quality of the conversion of the Gaussian mode to the Laguerre--Gaussian modes with OAM equal to $ l $ is quite low. This is due to differences in the spatial profiles of the driving field and quantum one. Laguerre--Gaussian modes with a nonzero momentum have a phase singularity at the point $ \rho = 0 $ and there is zero field intensity at this point, while for the Gaussian mode with $l = 0 $ we can say that the amplitude reaches a maximum at this point. The overlap of such dissimilar fields is small for any values of geometric parameters since in this case, the plane wave limit is not applicable. We cannot put the transverse size of a quantum mode much larger than the classical mode because it will inevitably lead to losses of information during the writing process.

The situation improves significantly when trying to increase and decrease the OAM of a quantum field with $ l\ne0 $. The Fig. \ref{FIGCHI2} shows the dependence of the coefficients $ \chi_{l, J}/ S_{l-J} $ for cases of writing by the driving field with the momentum  $ J = 1 $ (top) and $ J = -1 $ (bottom). This field configuration allows you to increase or decrease the OAM of the quantum field by 1.
\begin{figure}[h!]
\includegraphics[scale=0.4]{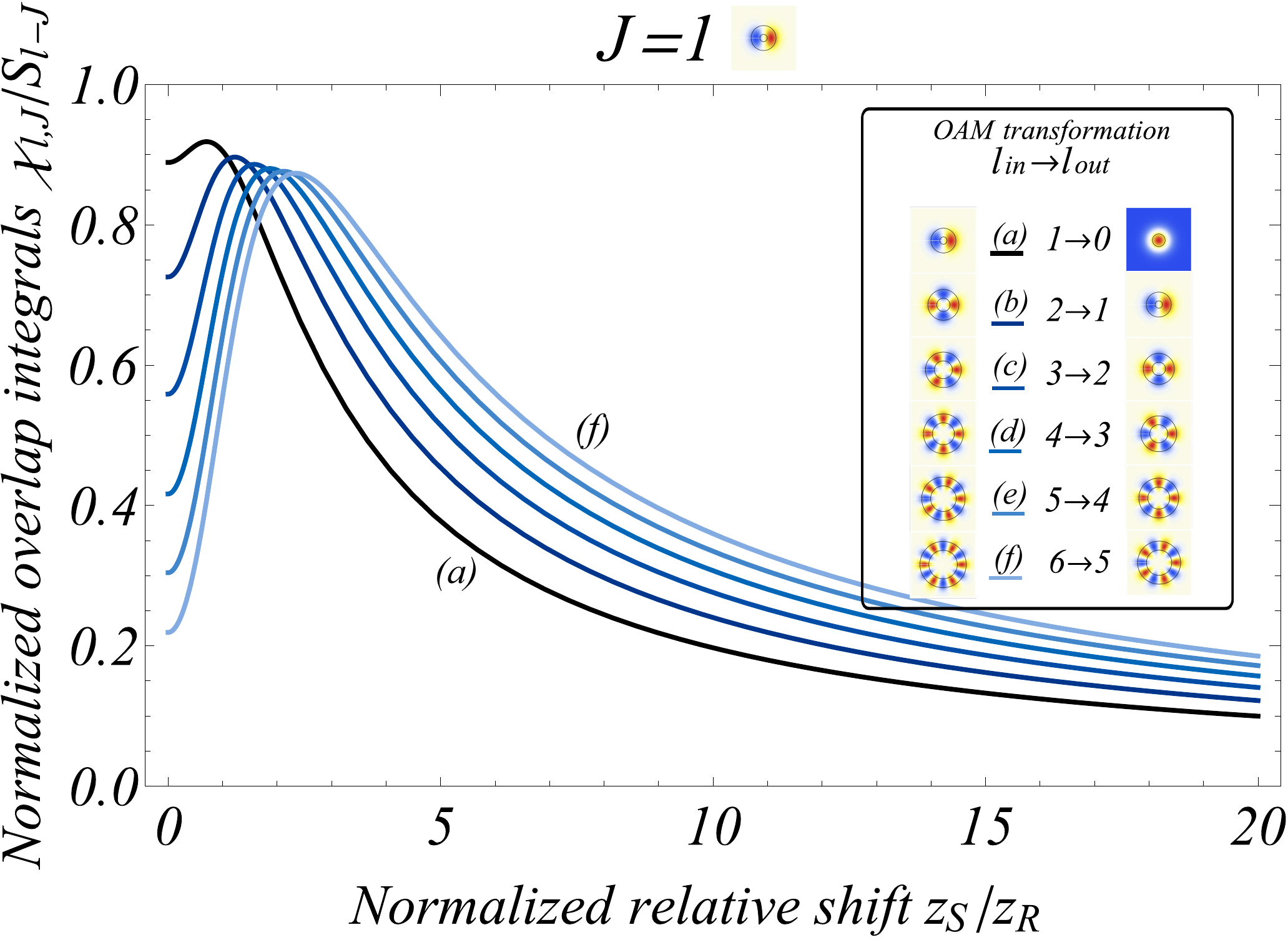}
\includegraphics[scale=0.4]{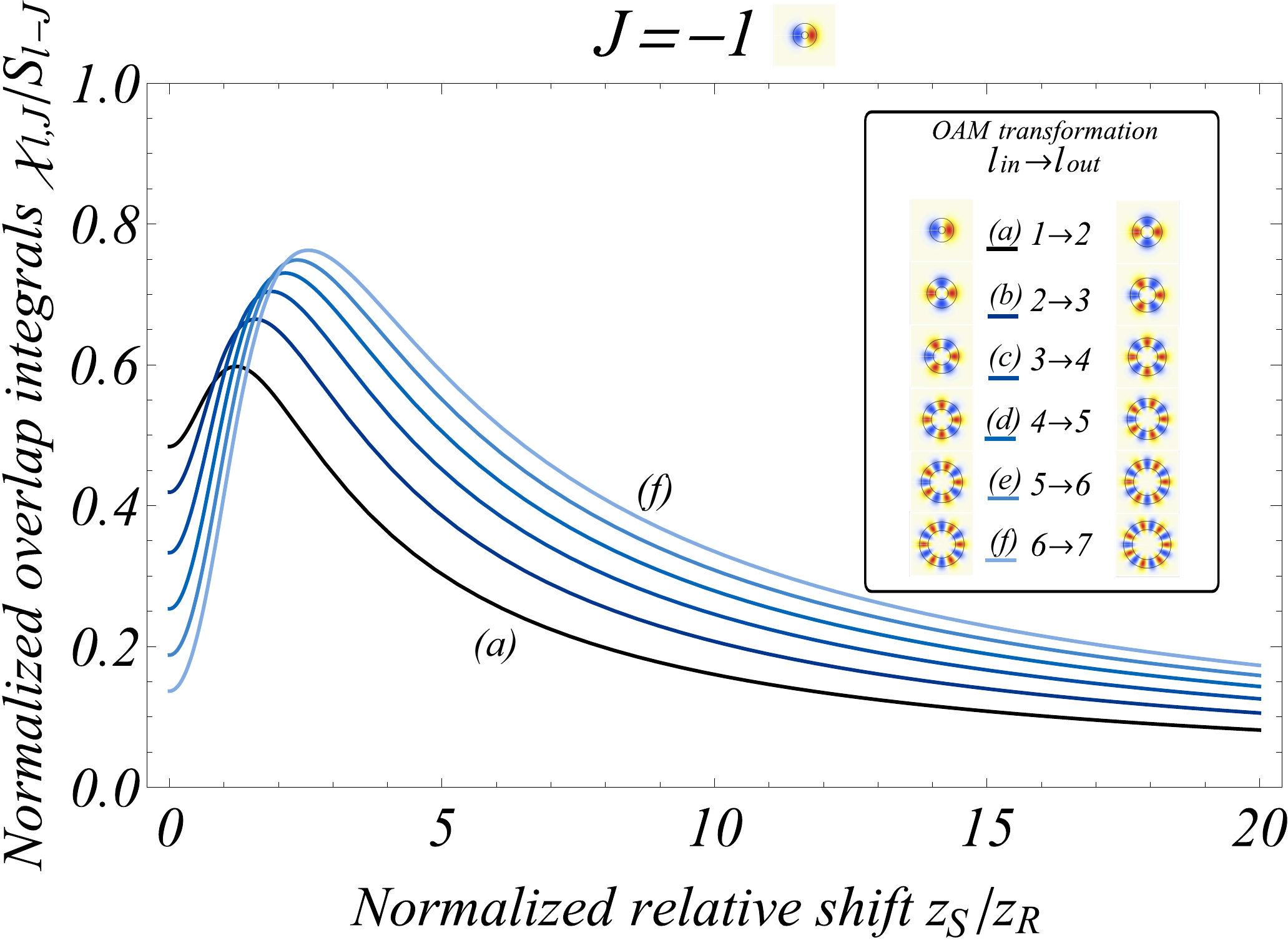}
\caption{The normalized coefficients $\chi_{l, J}/S_{l-J} $ depending on the normalized relative shift for the momentum of the control field $ J = 1 $ (top) and $ J = -1 $ (bottom).}
\L{FIGCHI2}
\end{figure}

It should be noted that each curve has a maximum corresponding to a certain value of $ z_S $, i.e. conversion of any angular momentum down or up by 1 can be carried out with high efficiency - about 0.9 for $-1$ conversion and $ 0.6-0.8 $ for $ + 1 $ conversion. The reason for this difference in the efficiencies of seemingly similar transformations is the effective "narrowing" of the coherence modes, which was discussed above in the same section. In this case, it is not so easy to determine the magnitude of the narrowing, since the Laguerre--Gaussian modes at the input of the cell have a spatial profile more complex than the Gaussian ones.

On the top of the Fig. \ref{FIGCHI2} the maximum value of the coefficients $ \chi_{l, J}/S_{lJ} $ decreases with increasing of $ l $ because the radius of the beam $ w_0 \sqrt {l}/2$ increases with $ l $  and in full accordance with all of the above, the effective transverse size of the restored mode is becoming smaller. But, since we are following the retrieved mode, the size of which is also smaller than the size of the one on the input ($ w_0 \sqrt{l-J}/2 <w_0 \sqrt{l}/2 $), the coefficients do not differ much from 1 for small values of $ l $. At large $ l $, the effect of lowering the efficiency from overlapping two different Laguerre-- Gaussian profiles becomes stronger than the effect of "focusing" \; of the flux of the number of photons — the integral flux through the cross-sectional area of the mode with the number $ l $ turns at the output into the flux through the cross-sectional area of the mode with a lower index $ l-J $ and therefore, a smaller area. It leads to an increase in the flux density and to greater efficiency.

On the bottom of Fig. \ref{FIGCHI2} we see the opposite situation: for small $ l $ values, the size of the retrieved mode exceeds both the size of the original and the effective size of the spin coherence discussed above, which is the reason for the low efficiency. As $ l $ grows, the negative effect decreases and the effective overlap of modes increases.

Conversion processes with large momentum $ J $ of the driving field also occur with high efficiency, however, they are similar to those considered above.

\section{Conclusion}
We developed a quantum memory protocol for modes with OAM. It turned out that neglection of the diffraction effects allow us to say that quantum field modes with different OAMs evolve independently of each other. The closed system of equations for the field operators $ \h a_l $ and the projections of collective spin coherences on the Laguerre--Gaussian profiles $ \h b_l $ and $ \h c_l $ could be constructed in this case.

By choosing various profiles of the driving field, one can simply store a quantum field with OAM with subsequent retrieving into the same mode or to another spatial mode with the different orbital momentum.

Storing and subsequent retrieval of the quantum states with OAM can be performed without additional (relative to standard Raman memory protocols) efficiency losses if writing and readout are carried out by a plane wave.

Varying the configuration of the driving fields at the writing and readout stages open up wide possibilities for converting the OAM of the quantum field simultaneously with storage. The appropriate set of the geometric parameters of the driving field ensures a high conversion quality (with a conversion coefficient of $ \approx 0.9 $).

In this case, the transformation does not depend on the direction of wave propagation during writing or readout stages.

The possibility of converting and simultaneously storing quantum states with OAM distinguishes the developed protocol from existing mode converters with OAM, for example, q-plates \cite{Slussarenko11} and phase holograms. The transformation using q-plates significantly depends on the intrinsic moment (rotation of the plane of polarization) of the converted light - only circularly twisted beams are converted and the intrinsic momentum, in the general case, is not preserved during such conversion \cite{Cardano12}. At the same time, our proposed method does not affect the polarization of light at the output of the cell.

  The main feature of the proposed conversion method is that it allows simultaneously storing quantum-statistical properties on a memory cell in one device and transferring these properties to modes with another OAM, which allows, in practice, to significantly reduce the number of optical elements in the experimental scheme.

  This work was supported by the RFBR (grants 19-32-90059, 19-02-00204, and 18-02-00648).

\appendix

\section{General solution of the Heisenberg equations and the kernel of integral transformations}\L{B}
The process of solving equations (\ref{HF1})--(\ref{HF3}) coincides with that described in \cite{Golubeva2012}, so here we present calculations only for the OAM conversion protocol described by the equations (\ref{HHF1})-- (\ref{HHF3}).
To solve the equations (\ref{HHF1})--(\ref{HHF3}), we perform the Laplace transform of the form:
\BY
f^s(z,s)=\int\limits_0^\infty dt f(z,t)e^{-st}.
\EY
Then the system of Heisenberg equations for Laplace-images of field and atomic variables is as follows:
\BY
&&\sqrt{S_l}\frac{d}{d z}\h a^s_l(z,s)=-g\sqrt{N} \sqrt{S_l}\h c^s_l(z,s)\L{L1}\\
&&\sqrt{S_l}(-\h c_l(z,0)+(s+i\Delta)\h c^s_l(z,s))=g \sqrt{N}\sqrt{S_l} \h a^s_l(z,s)\nn\\
&&+\Omega_m \frac{\chi_{l,m}}{\sqrt{S_l}\sqrt{S_{l-m}}}\h b^s_{l-m}(z,s)\sqrt{S_{l-m}}\;\;\;\;\;\;\;\;\\
&&\sqrt{S_{l-m}}(-\h b_{l-m}(z,0)+s\h b^s_{l-m}(z,s))=\nn\\
 &&=- \Omega_m \frac{\chi^*_{l,m}}{\sqrt{S_l}\sqrt{S_{l-m}}}\h c^s_l(z,s)\sqrt{S_l}.\L{L3}
\EY
For convenience, we redefine the variables:
\BY
&&\h{\td a}^s_l(z,s)=\sqrt{S_l}\h a^s_l(z,s);\;\;\;\h{\td c}^s_l(z,s)=\sqrt{S_l}\h c^s_l(z,s);\nn\\
&&\h{\td b}^s_{l-m}(z,s)=\sqrt{S_{l-m}}\h b^s_{l-m}(z,s);\nn\\
&&{\td \chi}_{l,m}=\frac{\chi_{l,m}}{\sqrt{S_l}\sqrt{S_{l-m}}}.
\EY
Since we will further discuss the processes of storing information  and its subsequent retrieving, for simplicity we will declare $c_l(z, 0) = 0 $, assuming that at the writing stage all atoms are prepared in the state $|1\> $ and there is no coherence between levels, and by the beginning of the readout stage, the coherence $ c_l (z, 0) $ is completely decays during storage. In this case, it is easy to obtain an equation describing the dynamics of a quantum field:
\BY
&&\frac{d}{dz}\h {\td a}^s_l(z,s)=-\Gamma^{l,m}_s\h {\td a}^s_l(z,s)-\alpha^{l,m}_s.
\EY
The following notation is introduced here:
\BY
&&\Gamma^{l,m}_s=\frac{g^2N}{2}\left(\frac{\mu_{l,m}}{s+i\widetilde{\Omega}_m|\td{\chi}_{l,m}|\mu_{l,m}}\right.\nn\\
&&\left.+\frac{\nu_{l,m}}{s-i\widetilde{\Omega}_m|\td{\chi}_{l,m}|\nu_{l,m}}\right), \\
&&\alpha_s^{l,m}(z,s) = \frac{1}{s^2+i\Delta s +\Omega^2_m|\td{\chi}_{l,m}|^2}\nn\\
&&\times\left( \Omega_m\td{\chi}_{l,m}\h{\td{b}}_{l-m}(z,0)\right),\;\;\;\;\;\;\\
&&\displaystyle\binom{\mu_{l,m}}{\nu_{l,m}}=1\pm\frac{r_{l,m}}{\sqrt{r_{l,m}^2+1}}=1\pm\frac{r}{\sqrt{r^2+|\td{\chi}_{l,m}|^2}},\\
&&\displaystyle\td{\chi}_{l,m}\widetilde{\Omega}_{m}=\td{\chi}_{l,m}\Omega_m\sqrt{r_{l,m}^2+1}\nn\\
&&r_{l,m}=\frac{\Delta}{2\Omega_m\td{\chi}_{l,m}}=\frac{r}{\td{\chi}_{l,m}}.\\
\EY
Here we can pass from operator quantities to c-numbers, assuming that further we will be interested only in normally ordered means. Then the solution of the equations (\ref{L1}) - (\ref{L3}) has the form:
\BY
&&{\td a}^s_l(z,s) = -g\sqrt{N} \int\limits_0^z dz^\prime \alpha_s^{l,m}(z^\prime,s)e^{-\Gamma^{l,m}_s(z-z^\prime)}\nn\\
&&+\td{ a}^s_l(0,s)e^{-\Gamma^{l,m}_s z}\\
&& {\td c}^s_l(z,s) = \frac{\Gamma^{l,m}_s}{g\sqrt{N}} \td{a}^s_l(z,s) + \alpha_s^{l,m}(z,s)\\
&&\td{b}^s_{l-m}(z,s) = \frac{1}{s}\left(\td{b}^s_{l-m}(z,0) - \Omega_m \td{\chi}_{l,m}\td{c}^s_l(z,s)\right).
\EY
We perform the inverse Laplace transform and write the solutions in the dimensionless variables $\tilde t = \Omega t $ and $ \tilde z = \displaystyle \frac{2g^2 N}{\Omega} z $. Assuming $ \h b^s_l(z, 0) = 0 $ at the writing stage, we obtain the solution for the “recorded” spin coherence $ b^W_l (\tilde {z}, \tilde {t}) $, returning to the original variables accepted in (\ref{L1} - \ref{L3}):
\BY
b^{W}_{l-m}(\tilde{z}, \tilde{t})&&=-\displaystyle\frac{g \sqrt{N}}{\Omega}\frac{\chi_{l,m}}{S_{l-m}}\int\limits_0^{\td{T}_W} d\tilde{t}^\prime a_l(0,\tilde{t}^\prime)G^{lm}_{ba}(\td{z},\tilde{t}-\tilde{t}^\prime).\;\;\;\;\;\;\;\;\;\;\;\;\;
\EY
The transformation kernel $G^{lm}_{ba} (\td{z}, \tilde{t} - \tilde {t}^\prime) $ can be written as follows:
\BY
&&G^{lm}_{ba}(\tilde{z},\td{t})=[f^{lm}_0(\tilde{z},\td{t},r)*f_0^{lm*}(\tilde{z},\td{t},-r)]=\nn\\
&&\int\limits_0^{\td{t}}d\td{t}^{\prime}f^{lm}_0(\tilde{z},\td{t}^\prime,r)f_0^{lm*}(\tilde{z},\td{t}-\td{t}^\prime,-r)\\
&&f^{lm}_0(\tilde{z},\td{t},r)=exp{\{-i\left(\sqrt{r^2+|\td{\chi}_{l,m}|^2} + r\right)\td{t}\}}\nn\\
&& \times J_0\left[\sqrt{\td{z}\td{t}\left(1+\frac{r}{\sqrt{r^2+|\td{\chi}_{l,m}|^2}}\right)}\right]\Theta(\td{t}).
\EY
Here $J_0$ is the Bessel function of the first kind, $\Theta(\td{t}) $ is the transmission function: $ \Theta(\td{t}) = 1; \; \; 0<\td{t}<\td{T} $, where $ \td{T} $ is the interaction time ($ \td{T}= \td{T^W} $ for writing and $\td{T} = \td{T^R } $ for readout).
We assume that during storage the coherence between the levels $ | 1> $ and $ | 2> $ is preserved without any losses ($ b^R_l (\tilde{z}, 0) = b^W_l (\tilde{z}, \td{T}_W $). Then for the readout stage we can write the following:
\BY
&&a^{R}_{l}(\tilde{L}, \tilde{t})= -\frac{\Omega}{g \sqrt{N}}\frac{\chi_{l,m}}{S_{l-m}}\int\limits_0^{\td{L}} d\tilde{z}\; b^W_{l-m}(\tilde{z}, \tilde{t}) G^{lm}_{ab}(\td{L}-\tilde{z}, \td{t}).\;\;\;\;\;\;\;\;\;
\EY
In the Raman limit $r\gg1 $, we can simplify the expressions above:
\BY
&&f^{lm}(\tilde{z},\td{t},r)\xrightarrow[r\gg1]{}\exp{\{-2i r\td{t}\}}J_0\left[\sqrt{2\td{z}\td{t}}\right]\Theta(\td{t}),\L{F1}\\
&&f^{lm}(\tilde{z},\td{t},-r)\xrightarrow[r\gg1]{}1.\L{F4}
\EY
For kernels in the Raman limit we can write:
\BY
G^{lm}_{ba}(\td{z},\td{t})=G^{lm}_{ab}(\td{z},\td{t})=\left[1\ast f(\td{z},\td{t},r)\right].
\EY

\bibliography{PaperQOAMT}


%

\end{document}